\documentclass[%
twocolumn,
superscriptaddress,
amsmath,amssymb,
aps,
]{revtex4-2}

\usepackage{hyperref}
\usepackage{amsmath}
\usepackage{amssymb}
\usepackage{graphicx}
\usepackage{color}
\usepackage{dcolumn}
\usepackage[export]{adjustbox}

\usepackage{tikz}	
\usepackage{pgfplots}
\usepackage{tikz-3dplot}

\usetikzlibrary{arrows,decorations.markings,calc,fadings,decorations.pathreplacing, patterns, decorations.pathmorphing, positioning,angles,quotes}

\begin{document}
\bibliographystyle{apsrev}

\title{Thin Pyramidal Cones in Nematic Liquid Crystal}

\author{Seyed Reza Seyednejad}
\affiliation{Institute for Advanced Studies in Basic Sciences, Zanjan 45195-159, Iran}

\author{Saeedeh Shoarinejad}
\affiliation{Department of Theoretical Physics and Nanophysics, Faculty of Physics, Alzahra University, Tehran, Iran}

\author{Mohammad Reza Mozaffari}
\email[Email: ]{m.mozaffari@qom.ac.ir}
\affiliation{Department of Physics, University of Qom, Qom 3716146611, Iran}

\author{Faezeh Amini Joneghani}
\affiliation{Department of Theoretical Physics and Nanophysics, Faculty of Physics, Alzahra University, Tehran, Iran}

\begin{abstract}
The present study investigates the arrangement of hollow pyramidal cone shells and their interactions with degenerate planar anchoring on the inner and outer surfaces of particles within the nematic host. The shell thickness is in order of the nematic coherence length. The numerical behavior of colloids is determined by minimizing the Landau-de Gennes free energy in the presence of the Fournier surface energy and using the finite element method. Colloidal pyramidal cones can orient parallel and perpendicular with the far director orientation. In the parallel alignment, we found the splay director distortion into the pyramid with two boojum defects at the inner and outer tips. The director shows bending distortion without defect patterns when the pyramid is aligned perpendicularly. They induce long-range dipolar interaction and can form nested structures in close contact. 
\end{abstract}

\date{\today}
\maketitle

\section{Introduction}
	The flat surfaces can form complex mathematical boundaries as nematic colloids with self-controllable defects in condensed matter physics~\cite{Park.PhysRevLett.117.2016}. The immersion of a colloidal particle in a uniform nematic liquid crystal (NLC) can cause a variety of elastic deviations in the nematic orientation around the particle~\cite{Poulin.science.275.1997}. The nematic orientation can appear as point or/and line singularities, known as defects, depending on the size and shape of the particle and the anchoring type of nematic molecules at the surface~\cite{Poulin.PhysRevE.57.1998}. In electrostatic analogy, the particle-defect pair induces elastic distortions as dipolar, quadrupolar, or hexadecapolar~\cite{Smalyukh.PhysRevLett.95.2005,Makoto.PhysRevLett.92.2004,Musevic.science.313.2006,Senyuk.ncomms.7.2016}. The particles and their accompanying defects exhibit anisotropic and long-range colloidal interactions in nematic media that depend on the individual Brownian motion of the particles~\cite{Araki.PhysRevLett.97.2006}. The spherical colloids cause quadrupolar elastic distortion symmetry as two surface boojums with degenerate planar anchoring~\cite{Smalyukh.PhysRevLett.95.2005, Mozaffari.SoftMatter.7.2011}. The dipolar and quadrupolar elastic distortion symmetries arise from the perpendicular anchoring on the spherical surfaces with the radial hedgehog and Saturn ring defects, respectively~\cite{Makoto.PhysRevLett.92.2004, Musevic.science.313.2006, Araki.PhysRevLett.97.2006}. Conical degenerate anchoring on a spherical particle disturbs the uniform nematic orientation as hexadecapolar symmetry with surface boojums and Saturn ring defects~\cite{Senyuk.ncomms.7.2016, dePablo.ncomms.10.2019, Seyednejad.PhysRevE.99.2019}.
	
	Using novel techniques to produce complex colloidal particles with different topological genus numbers in the NLC has extended defect studies from a theoretical perspective~\cite{Senyuk.nature.493.2013,Yuan.naturematerial.17.2018,Hashemi.NatureCommunications.8.2017, Lapointe.science.326.2009,Senyuk.PhysRevE.91.2015,Seyednejad.PhysRevE.98.2018,Yuan.nature.570.2019}. Compared to spherical particles, the elastic deformations around the geometries with low symmetry depend critically on their shape and spatial arrangement in the nematic host. Handle-body colloids with a certain genus number and perpendicular surface anchoring can have different point and ring defect arrangements in the uniform NLC. Depending on the handle-body particles, the individual singularities are predictable by the genus number~\cite{Senyuk.nature.493.2013}. The elastic distortions of chiral superstructure colloids with strong surface boundary conditions can lead to local chiral configurations in the nematic field. The effects of the distortions give rise to anisotropic and long-range interactions between like- and opposite-handed particles with different geometric parameters~\cite{Yuan.naturematerial.17.2018}. The iteration of Koch fractals provides the conditions to study induced topological singularities in the nematic orientation. The number of topological defect pairs increases exponentially in the fractal iteration~\cite{Hashemi.NatureCommunications.8.2017}. The number of sides of polygonal platelets in the NLC gives the dipolar and quadrupolar elastic symmetries. They can be employed to design new composite colloidal structures~\cite{Lapointe.science.326.2009}. The pentagonal platelets form quasi-crystalline colloidal lattices as Penrose tiling patterns in NLC layers. The quadrupolar platelets are more energetically optimal than dipolar platelets in close contact interactions. The interactions of polygonal truncated pyramids depend geometrically on the tile fragments and arrays with long-range dipolar interactions~\cite{Senyuk.PhysRevE.91.2015,Seyednejad.PhysRevE.98.2018}. They attract (repel) each other in the parallel (antiparallel) alignments at the base-base configurations. In the side-side cases, the particles attract (repel) each other in the antiparallel (parallel) alignments. The thin hexagonal prisms can induce elastic monopole moments, which balance the elastic torques with optical torques. The repulsive and attractive interactions in elastic monopoles act differently than in electrostatics~\cite{Yuan.nature.570.2019}. A colloidal particle with homeotropic (planar) anchoring can be elastically trapped on the tip of a pyramid with planar (homeotropic) anchoring. The trapping potentials have a short-range effect and show good resilience against thermal fluctuations~\cite{Silvestre.PhysRevLett.112.2014}. The substrate patterns of pyramidlike protrusions can give rise to the $2$D and $3$D self-assembly of nematic colloidal structures.
		
	The flat surfaces, known as without boundary geometries, do not contain defects in uniform nematic media. When these surfaces have formed into hollow pyramidal cones, the elastic deformations induce dipolar anisotropic interactions through two fractional boojum defects of opposite signs near the inner and outer vertices of the pyramids. Experimental observations show parallel and perpendicular alignments of the apex axis with the nematic orientation in the far field. In comparison with the non-spherical colloids, the formed flat surfaces can be morphed to complex shapes with self-compensating defects. They can form nested configurations in both parallel and perpendicular arrangements~\cite{Park.PhysRevLett.117.2016}. 

	Here, motivated by the experimental finding of colloidal pyramidal cones (CPCs)~\cite{Park.PhysRevLett.117.2016}, we have proposed a simple approach to specify the nematic field at the boundaries of thin surfaces. The nematic field plays a critical role at these boundaries to match both side surface director anchorings and around bulk orientations when the defects are absent. We first numerically investigate the equilibrium arrangements of a single colloidal pyramidal cone with degenerate planar anchoring in the NLC host. Then, the colloidal repulsive and attractive interactions are studied to compare experimental results in self-assemblies and nested configurations~\cite{Park.PhysRevLett.117.2016}. We use the finite element method to obtain the orientation of the nematic equilibrium field by minimizing Landau-de Gennes's free energy in the present Fournier and Galatola surface energy.
	
	\begin{figure}[t]
		\centering
		\includegraphics[scale=0.425]{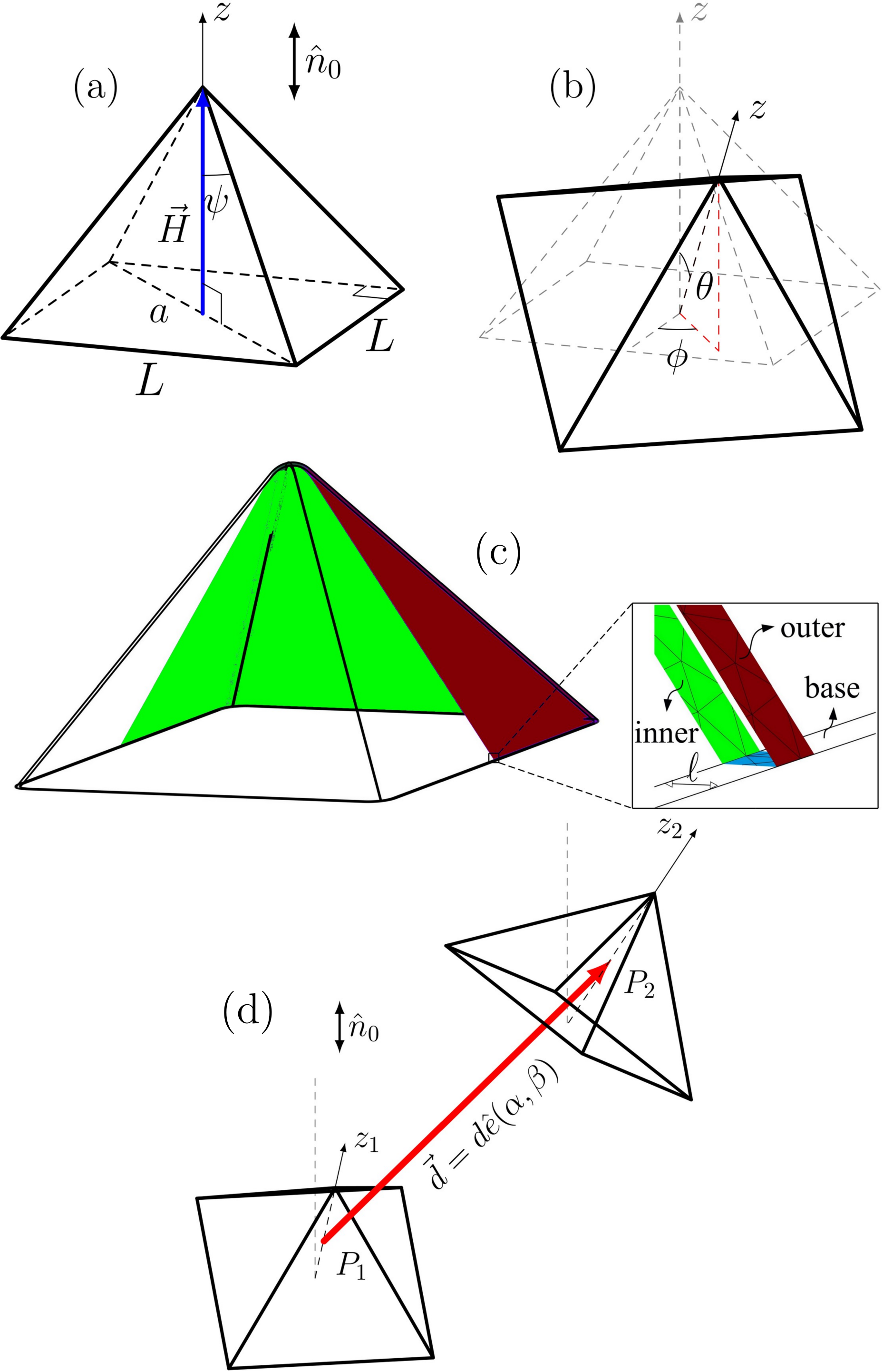}
		\caption{(a) Schematic representation of the colloidal pyramidal cone skeleton with the relevant parameters: side length of the square base $L$, apex angle $2\psi$, height from base to apex $H(=|\vec{H}|)$, diagonal length of the square base $a (= \sqrt{2}L = 2H\tan\psi)$ and far-nematic orientation $\hat{n}_0$. (b) Schematic representation of the arbitrary alignment of the pyramidal axis around $\hat{n}_0$ with polar ($\theta$) and azimuthal ($\phi$) angles. (c) The meshing of side faces with two independent inner and outer boundaries as a thin pyramidal shell where $\ell$ is the thickness of the shell. The vertices on the narrow plane in the shell base are part of the principal mesh. Curve the apex and edges on both sides of the pyramidal shell sides. (d) Schematic representation of two identical pyramidal cones. The particle-particle distance defines as $P_1P_2 = |\vec{d}|$, where $P_1$ and $P_2$ are the midpoints of the heights.} 
		\label{fig01}
	\end{figure}

\section{Numerical method}
	Fig.~\ref{fig01}(a) schematically shows the CPC geometry. We have chosen the diagonal length of the square pyramid base as the system length scale ($a = 1\:\mu\text{m}$). $\vec{H}$ is a vector from center of the base to apex. As shown in Fig.~\ref{fig01}(b), the spatial CPC alignment is accounted for by $\theta$ and $\phi$ angles. We used a thin pyramid shell mesh, as shown in Fig.~\ref{fig01}(c), to solve the numerical limitations of applying boundary conditions on both the inside and outside of the hollow pyramid. We fixed the thickness of the shell in the order of the nematic coherence length. Fig.~\ref{fig01}(d) shows schematically two identical CPCs. They are located spatially in the center of the nematic cell ($L_x = L_y = L_z = 15a$) with the distance $d$ between the particles and the corresponding spherical angles, $\alpha$, and $\beta$.
	
	Here we locally use a symmetric and traceless second rank tensor, $Q_{ij}=S(3\hat{n}_i\hat{n}_j-\delta_{ij})/2+P(\hat{l}_i\hat{l}_j-\hat{m}_i\hat{m}_j)/2$, with five independent components to study the nematic fluid. The largest eigenvalue and the corresponding eigenvector of the nematic tensor are the scalar order parameter, $S$, and the director orientation $\hat{n}$, respectively~\cite{deGennes.95}. $\hat{l}$ and $\hat{m}$ are the two smaller eigenvectors, and $P$, so called biaxiality order parameter, determined by the difference of the second and third large eigenvalues, equilibrium value of which is zero mostly, due to the uniaxial nature of this study, except near the defects. Due to the presence of CPC particles in the uniform nematic host with planar degenerate anchoring, the director undergoes elastic distortions around the particles. The Landau-de Gennes free energy can explain such distortions in terms of the tensor order parameter and its spatial derivatives as follows,
	
	\begin{eqnarray}
	\begin{split}
	\mathcal{F}_\mathrm{LdG} = \int_\Omega\mathrm{d}\mathcal{V}&\left[\frac{a_0(T-T^\ast)}{2}Q_{ij}Q_{ji}-\frac{B}{3}Q_{ij}Q_{jk}Q_{ki}\right.\\
	&\left.\quad+\frac{C}{4}(Q_{ij}Q_{ji})^2  +\frac{L_1}{2}(\partial_kQ_{ij})^2\right],
	\end{split}
	\label{eq1}
	\end{eqnarray}
	where the indices refer to Cartesian coordinates, and the Einstein summation convention is assumed. $\Omega$ is the volume occupied by the NLC~\cite{Kleman.03}. The first three terms generally specify the equilibrium scalar order parameter, $S_\text{eq}$, in terms of temperature. The positive coefficients $a_0$, $B$, and $C$ are material-dependent and temperature independent. $T^\ast$ is the nematic supercooling temperature. The last term determines the contribution of elastic distortions in the nematic media. $L_1$ is related to the Frank elastic constants in one-constant elastic approximation as $K_\text{splay} = K_\text{twist} = K_\text{bend} = 9L_1S^2_\text{eq}/2$, where $S_\text{eq} = (B/6C)(1+\sqrt{1-24a_0(T-T^\star)C/B^2})$. The spatial structure of defects is represented by using isosurfaces of scalar order parameter with $S \le S_\text{eq}/2$.
	
	Fournier and Galatola introduced the surface energy for planar degenerate anchoring as,
	
	\begin{eqnarray}
	\begin{split}
	\mathcal{F}_\text{S} &= W_1\int_{\partial\Omega}\mathrm{d}\mathcal{S}(\tilde{Q}_{ij} -\tilde{Q}_{ij}^\perp)(\tilde{Q}_{ji} -\tilde{Q}_{ji}^\perp)\\
	&+W_2\int_{\partial\Omega}\mathrm{d}\mathcal{S}(\tilde{Q}_{ij}^2 - (3S_\text{eq}/2)^2)^2,
	\end{split}
	\label{eq2}
	\end{eqnarray}
	where $\tilde{Q}_{ij}^\perp = (\delta_{ik} - \hat{\nu}_{i}\hat{\nu}_{k})\tilde{Q}_{kl}(\delta_{kj} -\hat{\nu}_{k}\hat{\nu}_{j})$ is the projection of $\tilde{Q}_{ij} = Q_{ij} + S\delta_{ij}/2$ onto the both side surfaces of the CPC particles. $W_1>0$ and $W_2>0$ are surface anchoring strengths, and $\hat{\nu}$ is the normal to the surface. $\partial\Omega$ implies all inner and outer surfaces on the CPCs. The first term forces the director to lie in the plane defined by $\hat{\nu}$. The second term ensures that the trace of $\tilde{Q}_{ij}^2$ must be equal to $(3S_\text{eq}/2)^2$~\cite{Fournier.EurophysLett.72.2005}. According to the past studies, here we relax the second term to evaluate the variation of the scalar order parameter on the surface~\cite{Vilfan.PhysRevLett.101.08,Mozaffari.SoftMatter.7.2011}.

	\begin{figure}[b]
		\centering
		\includegraphics[scale=0.70]{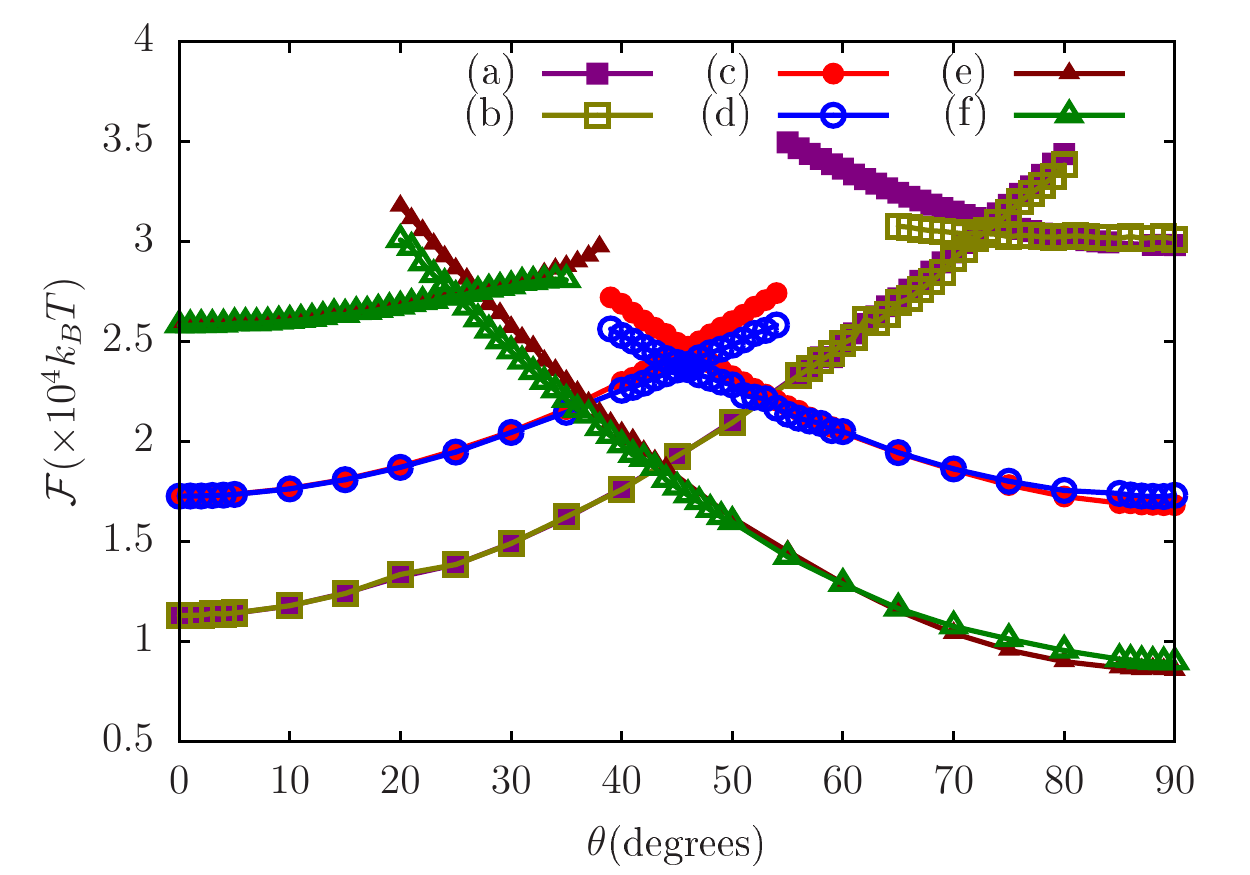}
		\caption{The energy of an isolated CPC in a uniform nematic media as a function of polar angles at different apex and azimuth angles, (a) $2\psi = 70^\circ$ and $\phi=0$, (b) $2\psi = 70^\circ$ and $\phi=45^\circ$, (c) $2\psi = 95^\circ$ and $\phi=0^\circ$, (d) $2\psi = 95^\circ$ and $\phi=45^\circ$, (e) $2\psi = 120^\circ$ and $\phi=0^\circ$, and (f) $2\psi = 120^\circ$ and $\phi=45^\circ$.}
		\label{fig02}
	\end{figure}

	\begin{figure}[t]
		\centering
		\includegraphics[scale=0.425]{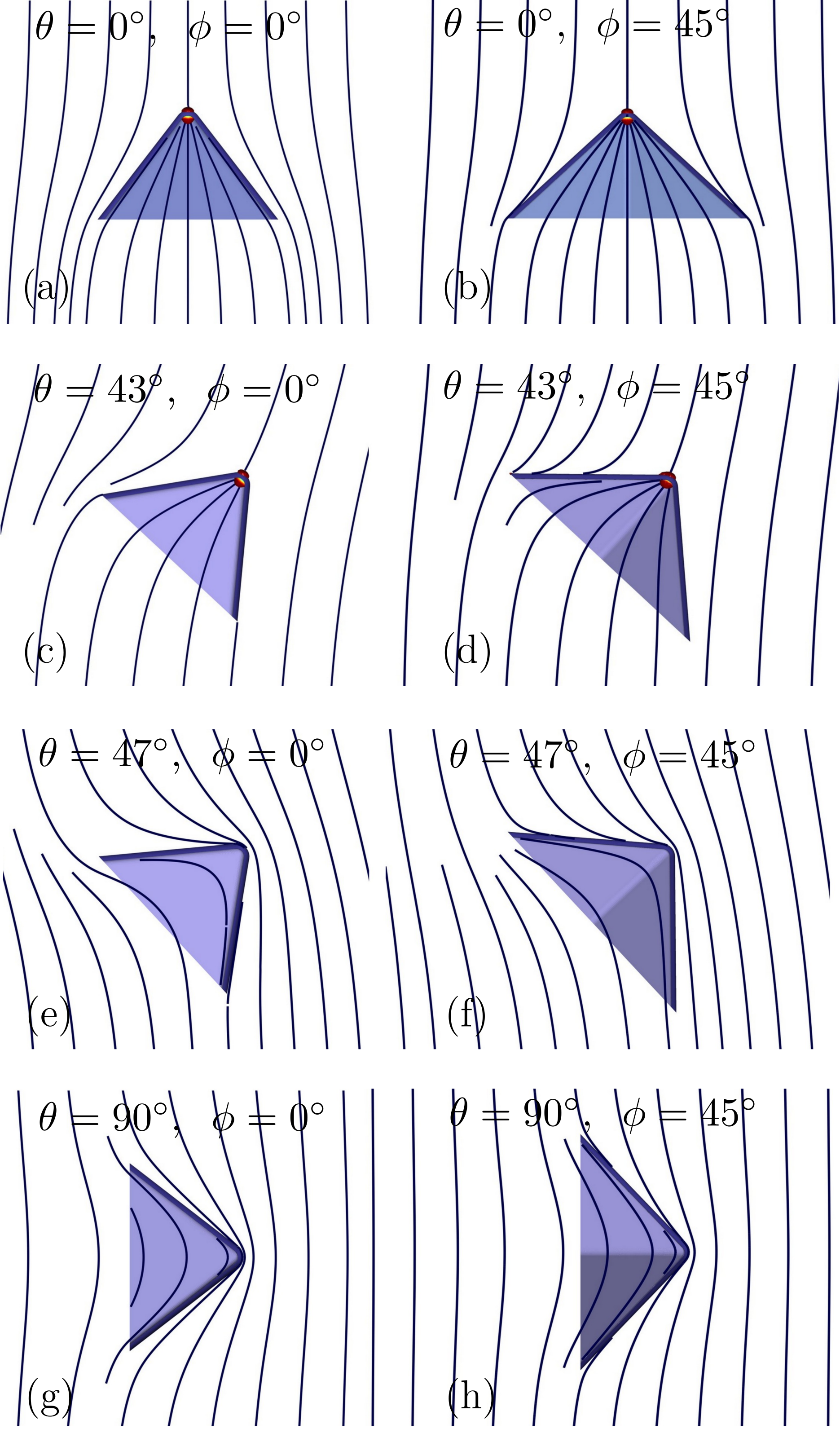}
		\caption{The equilibrium director orientations and boojum defects around the CPC with $2\psi = 95^\circ$ in the uniform nematic media, (a) $\theta = 0^\circ$ and $\phi = 0^\circ$, (b) $\theta = 0^\circ$ and $\phi = 45^\circ$, (c) $\theta = 43^\circ$ and $\phi = 0^\circ$, (d) $\theta = 43^\circ$ and $\phi = 45^\circ$, (e) $\theta = 47^\circ$ and $\phi = 0^\circ$, (f) $\theta = 47^\circ$ and $\phi = 45^\circ$, (g) $\theta = 90^\circ$ and $\phi = 0^\circ$, and (h) $\theta = 90^\circ$ and $\phi = 45^\circ$.}
		\label{fig03}
	\end{figure} 

	We use the 5CB parameters as $a_0 = 0.087\times 10^6\;\mathrm{J}/\mathrm{m}^3\mathrm{K}$, $T^\ast = 307.15\;\mathrm{K}$, $T = 305.17\;\mathrm{K}$, $B = 2.12\times 10^6\;\mathrm{J}/\mathrm{m}^3$, $C = 1.73\times 10^6\;\mathrm{J}/\mathrm{m}^3$, $L_1 = 4\times 10^{-11}\;\mathrm{J}/\mathrm{m}$~\cite{Kralj.PhysRevA.43.91}. The anchoring strength is assumed to be sufficiently large with $W_1 =10^{-2}\;\mathrm{J}/\mathrm{m}^2$.

	To minimize the total free energy ($\mathcal{F} = \mathcal{F}_\text{LdG} + \mathcal{F}_\text{S}$)~\cite{Mozaffari.SoftMatter.7.2011}, we have employed a finite element method. We have decomposed the $3$D domain into tetrahedral elements with the automatic mesh generator Gmsh. The tensor order parameter components have the first order interpolation within the element~\cite{Gmsh}. The mesh sizes control the accuracy of the numerical derivatives. An iterative conjugate gradient method uses to perform the minimization process~\cite{Press.92}. For each configuration, the minimization procedure was stopped when the relative free energy improvements dropped below $10^{-6}$.

	\begin{figure*}[t]
		\centering
		\includegraphics[scale=0.71]{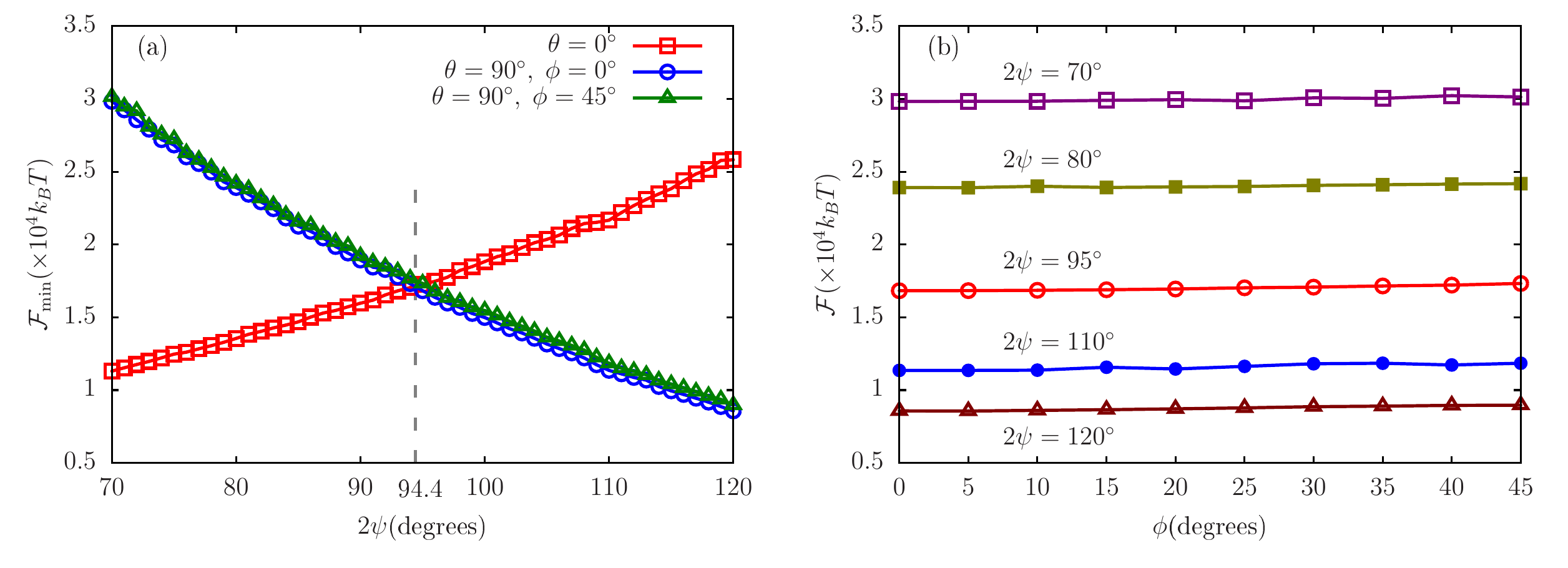}
		\caption{(a) The minimum energy value of an isolated CPC in a uniform nematic media as a function of apex angle at $\theta=0^\circ$ and $\theta=90^\circ$. The azimuth dependence of the minimum energy is considered in $\theta=90^\circ$ for $\phi=0^\circ$ and $45^\circ$. (b) The energy dependence of azimuth angles in $\theta=90^\circ$ at different apex angles.}
		\label{fig04}
	\end{figure*}

\section{Results and Discussion}

	In Fig.~\ref{fig02}, the spatial arrangements of a single CPC at three apex angles ($2\psi=70^\circ$, $95^\circ$, and $120^\circ$) have been considered energetically in a uniform nematic media. The cross energies result from the coincidence of two different energy branches at each apex angle. The energy curves on the left side of the crossing points are independent of the azimuthal arrangements of the CPCs, but they show some differences at $\phi=0^\circ$ and $45^\circ$ on the right side. Each energy curve shows an attractive elastic torque with a minimum value at $\theta=0^\circ$ or $90^\circ$. For $2\psi<95^\circ$, the equilibrium spatial arrangement is $\theta=0^\circ$. On the contrary, $\theta=90^\circ$ is the global minimum for $2\psi>95^\circ$. Our calculations show that $2\psi\simeq 95^\circ$ has the same minimum energy values at $\theta=0^\circ$ and $90^\circ$. Here, we have focused our studies on the CPCs with the apex angle of $95^\circ$.	

	In Fig.~\ref{fig03}, we have displayed the equilibrium director orientations and the defects for $2\psi=95^\circ$. The director field shows two different arrangements around the cross energy in Fig.~\ref{fig02}($\theta_c=45^\circ$). 
	
	In $\theta<45^\circ$, two boojum defects appear at the inner and outer pyramid tips. From Figs.~\ref{fig03}(a)-(d), it is easy to realize that the splay distortion is the dominant elastic deformation within the CPC. As shown in Figs.~\ref{fig03}(e)-(h), our calculations show no defects at the inner or outer pyramid tip for $\theta>45^\circ$, which is not consistent with experimental observations. Here the elastic distortions show a bending deformation within the pyramid. Indeed, the left and right sides of cross energies in Fig.~\ref{fig02} show two different director structures. We have obtained the same results for the other apex angles ($70^\circ\le 2\psi \le 120^\circ$).
	
	Figure \ref{fig04}(a) gives a better insight into the minimum energies in $\theta=0^\circ$ and $90^\circ$ at different apex angles. The plot shows a critical behavior around $95^\circ$ where $\theta=0^\circ$ and $90^\circ$ have the same minimum energy at the same apex angle. We have found that the splay and bend elastic distortions are the governing elastic deformations on the left and right sides, respectively. The plot behaves like a phase diagram for selective spatial arrangements of CPCs in the nematic host. The results are independent of the azimuthal directions of the CPCs at $\theta=0^\circ$. For $\theta=90^\circ$, the energies in $\phi=0^\circ$ and $45^\circ$ show the same trend with a small difference.In Fig.~\ref{fig04}(b), we consider the energy of the CPC orientation as a function of the azimuth angle $\phi$ in $\theta=90^\circ$. The nearly straight lines show small positive slopes for each apex angle. We can look in detail for increasing energies from below with decreasing CPC apex angle.
	
	\begin{figure*}[t]
		\centering
		\includegraphics[scale=0.55]{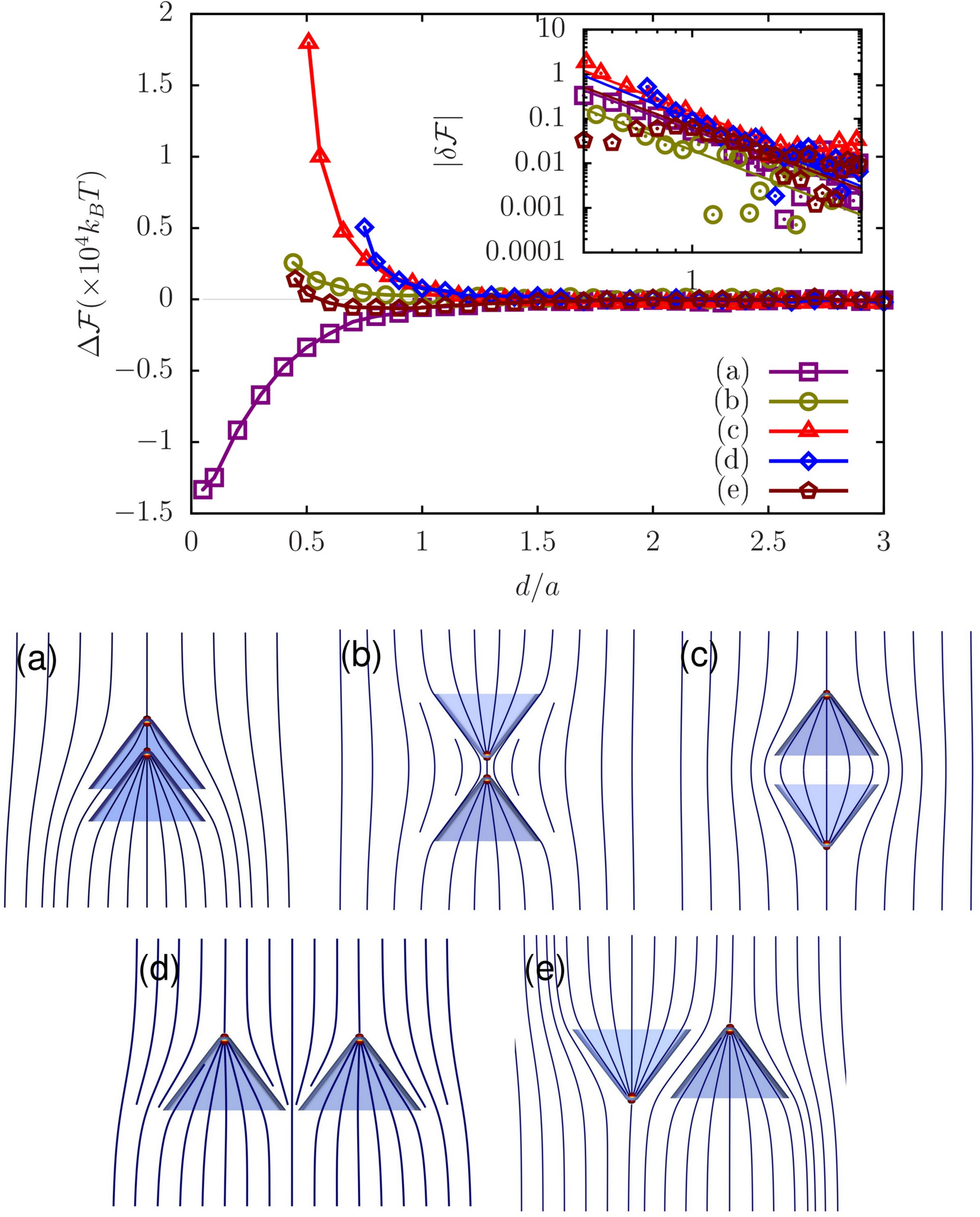}
		\caption{Interactions of two identical CPCs with $2\psi=95^\circ$ and $\vec{H}\parallel\hat{n}_0$ in terms of particle-particle distance, $d$, where $\Delta\mathcal{F} = \mathcal{F} - 2\mathcal{F}_\text{iso}$ and $\mathcal{F}_\text{iso}(\theta=0^\circ)=1.728\times10^4k_\text{B}T$. The particle-particle arrangements, (a) coaxial and parallel ($\alpha=0^\circ$ and $\theta_1=\theta_2 = 0^\circ$), (b) coaxial and apex-apex antiparallel ($\alpha=0^\circ$, $\theta_1=0$ and $\theta_2 = 180^\circ$), (c) coaxial and base-base antiparallel ($\alpha=0^\circ$, $\theta_1=180^\circ$ and $\theta_2 = 0^\circ$), (d) non-coaxial and parallel ($\alpha=90^\circ$, $\theta_1=0$ and $\theta_2 = 180^\circ$), and (e) non-coaxial and antiparallel ($\alpha=90^\circ$, $\theta_1=180^\circ$ and $\theta_2 = 0^\circ$). The inset log-log plot shows the fitting of the numerical results by a power law function as $c_0 + c_1/(d/R)^{c_2}$ that $\delta\mathcal{F} = \mathcal{F} - c_0$. The unit of $c_0$ and $c_1$ is $10^4 k_B T$. The lines correspond to (a) $c_0=3.445\pm0.002$, $c_1=-0.056\pm0.002$ and $c_2=3.0\pm0.1$, (b) $c_0=3.456\pm0.003$, $c_1=0.020\pm0.002$ and $c_2=3.1\pm0.2$, (c) $c_0=3.408\pm0.003$, $c_1=0.140\pm0.004$ and $c_2=3.1\pm0.1$, (d) $c_0=3.435\pm0.002$, $c_1=0.098\pm0.005$ and $c_2=3.2\pm0.2$, and (e) $c_0=3.453\pm0.002$, $c_1=-0.065\pm0.005$ and $c_2=3.0\pm0.2$. The distances between particles are as in (a) $0.20a$, (b) $0.54a$, (c) $0.66a$, (d) $0.80a$, and (e) $0.60a$.}
		\label{fig05}
	\end{figure*}

	From now on, we focus on the interaction between two identical CPCs with the apex angle of $95^\circ$ and equilibrium polar orientations $0^\circ$ and $90^\circ$, which energetically have the same magnitude with different elastic deformations, as mentioned above.
	
	Figure \ref{fig05} shows the effective interaction potential between two CPCs with $\vec{H}\parallel\hat{n}_0$ in terms of inter-particle distances for the selected particle-particle arrangements as shown in Figs.~\ref{fig05}(a)-(e). Particles in coaxial and parallel (Fig.~\ref{fig05}(a)) and non-coaxial and antiparallel (Fig.~\ref{fig05}(e)) arrangements attract each other. In the other configurations, coaxial and apex-apex antiparallel (Fig.~\ref{fig05}(b)), coaxial and base-base antiparallel (Fig.~\ref{fig05}(c)), and non-coaxial and parallel (Fig.~\ref{fig05}(d)), the CPCs repel each other.
	
	Regardless of the distances between the particles, all attractive and repulsive interactions are comparable to dipole-dipole interaction, where the height vector of the pyramids behaves like a dipole moment~\cite{jackson}. Here, the inner and outer boojums of the apexes cause dipole elastic distortions that can illustrate repulsive and attractive interactions. We have described the dependence of the interaction energies between the particles by a power law function. The inset log-log plot shows that the long-range interactions are nearly proportional to $\sim(d/R)^{-3}$, which is consistent with experimental observations~\cite{Park.PhysRevLett.117.2016}. The nested CPCs (in Fig.~\ref{fig05}(a)) show an energetically favorable structure in agreement with experimental observations. These structures reduce the elastic deformations between adjacent surfaces~\cite{Park.PhysRevLett.117.2016}. Due to small thermal fluctuations and spontaneous symmetry breaking, the spatial arrangement of CPCs with repulsive interaction in Fig.~\ref{fig05}(b) can transform into attractive interaction in Fig.~\ref{fig05}(e). In Fig.~\ref{fig05}(c), the repulsive interactions show that the formation of octahedral shapes seems almost impossible. In Figs.~\ref{fig05}(b) and \ref{fig05}(c), the interactions are independent of azimuth angle differences (see Supplemental Material Fig. S1~\cite{Supplemental}). Fig.\ref{fig05}(e) shows that the CPCs can locally form wavy surfaces in the nematic host to control spherical colloidal structures~\cite{Luo.ncomms.9.2018}. In Figs.\ref{fig05}(b)-(d) with repulsive interactions, the director shows mirror symmetry about the plane passing through the bisector of the distances between two particles.

	\begin{figure*}[t]
		\centering
		\includegraphics[scale=0.77]{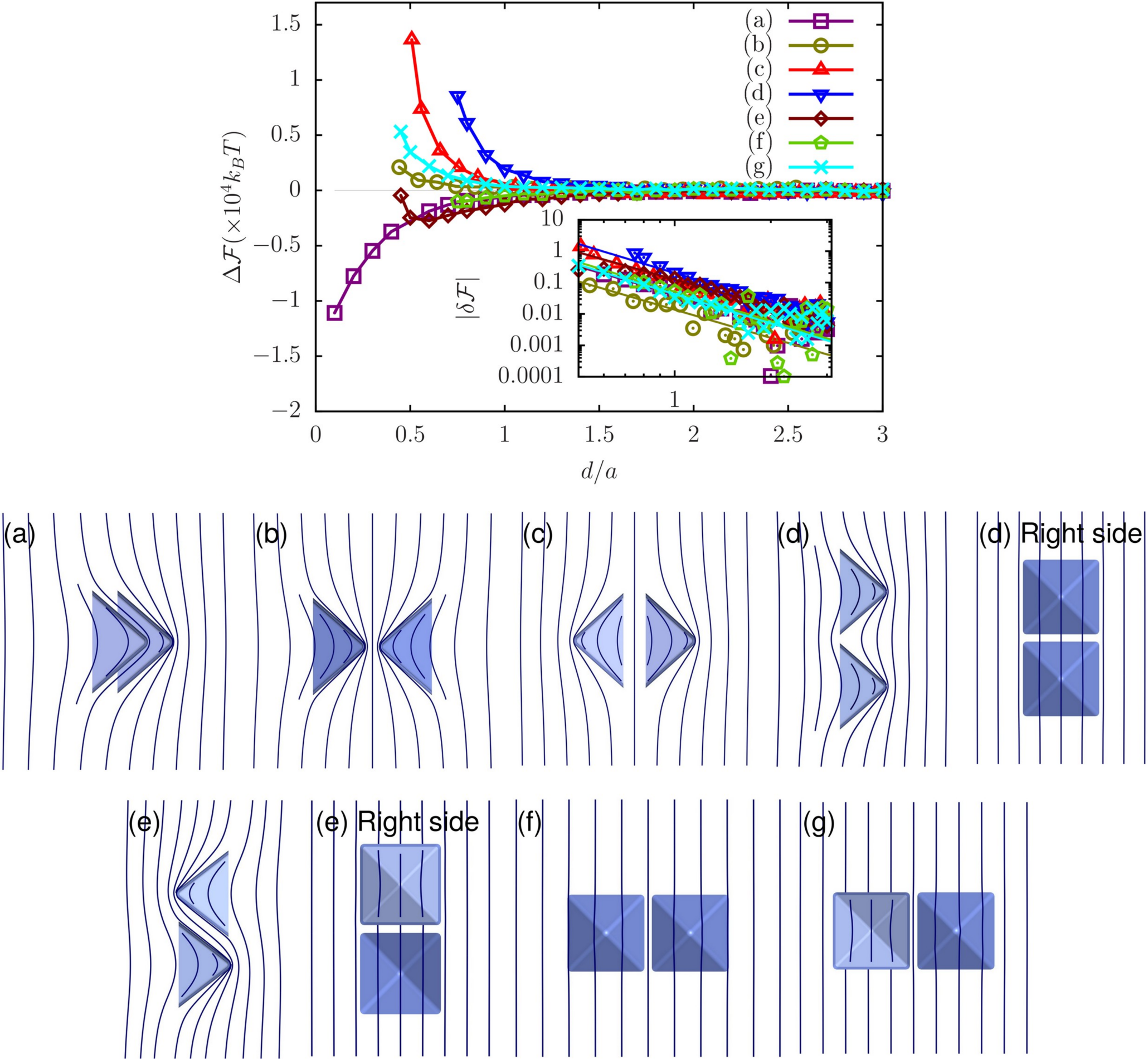}
		\caption{Interactions of two identical CPCs with $2\psi=95^\circ$ and $\vec{H}\perp\hat{n}_0$ in terms of particle-particle separations, $d$, where $\Delta\mathcal{F} = \mathcal{F} - 2\mathcal{F}_\text{iso}$ and $\mathcal{F}_\text{iso}(\theta=90^\circ, \phi=0^\circ)=1.682\times10^4k_\text{B}T$. The particle arrangements, (a) coaxial and parallel ($\alpha=90^\circ$, $\theta_1=\theta_2 = 90^\circ$ and $\phi_1=\phi_2=0^\circ$), (b) coaxial and apex-apex antiparallel ($\alpha=90^\circ$, $\theta_1=90^\circ$, $\theta_2=-90^\circ$ and $\phi_1=\phi_2=0^\circ$), (c) coaxial and base-base antiparallel ($\alpha=90^\circ$, $\theta_1=-90^\circ$, $\theta_2 = 90^\circ$ and $\phi_1=\phi_2=0^\circ$), (d) noncoaxial and parallel ($\alpha=0^\circ$, $\theta_1=\theta_2 = 90^\circ$ and $\phi_1=\phi_2=0^\circ$) (the figure on the right is for better illustration), (e) noncoaxial and antiparallel ($\alpha=0^\circ$, $\theta_1=90^\circ$, $\theta_2 =-90^\circ$ and $\phi_1=\phi_2=0^\circ$) (the figure on the right is for better illustration), (f) noncoaxial and parallel ($\alpha=90^\circ$, $\theta_1=\theta_2 =90^\circ$, and $\phi_1=\phi_2=0^\circ$) and (g) noncoaxial and antiparallel ($\alpha=90^\circ$, $\theta_1=-90^\circ$, $\theta_2 = 90^\circ$ and $\phi_1=\phi_2=0^\circ$). The inset log-log plot shows the fitting of the numerical results by a power law function as $c_0 + c_1/(d/R)^{c_2}$ that $\delta\mathcal{F} = \mathcal{F} - c_0$. The unit of $c_0$ and $c_1$ is $10^4 k_B T$. The lines correspond to (a) $c_0=3.354\pm0.002$, $c_1=-0.042\pm0.001$ and $c_2=2.9\pm0.1$, (b) $c_0=3.371\pm0.002$, $c_1=0.014\pm0.001$ and $c_2=3.0\pm0.1$, (c) $c_0=3.321\pm0.003$, $c_1=0.110\pm0.003$ and $c_2=3.1\pm0.1$, (d) $c_0=3.341\pm0.003$, $c_1=0.202\pm0.008$ and $c_2=3.1\pm0.1$, (e) $c_0=3.366\pm0.002$, $c_1=-0.125\pm0.004$ and $c_2=2.9\pm0.1$, (f) $c_0=3.365\pm0.003$, $c_1=-0.051\pm0.004$ and $c_2=3.1\pm0.3$, and (g) $c_0=3.364\pm0.002$, $c_1=-0.043\pm0.002$ and $c_2=3.1\pm0.1$. The distances between particles are as in (a) $0.20a$, (b) $0.54a$, (c) $0.66a$, (d) $0.80a$, (e) $0.60a$, (f) $0.80a$ and (g) $0.80a$.}
		\label{fig06}
	\end{figure*}
	
	\begin{figure*}[t]
		\centering
		\includegraphics[scale=0.64]{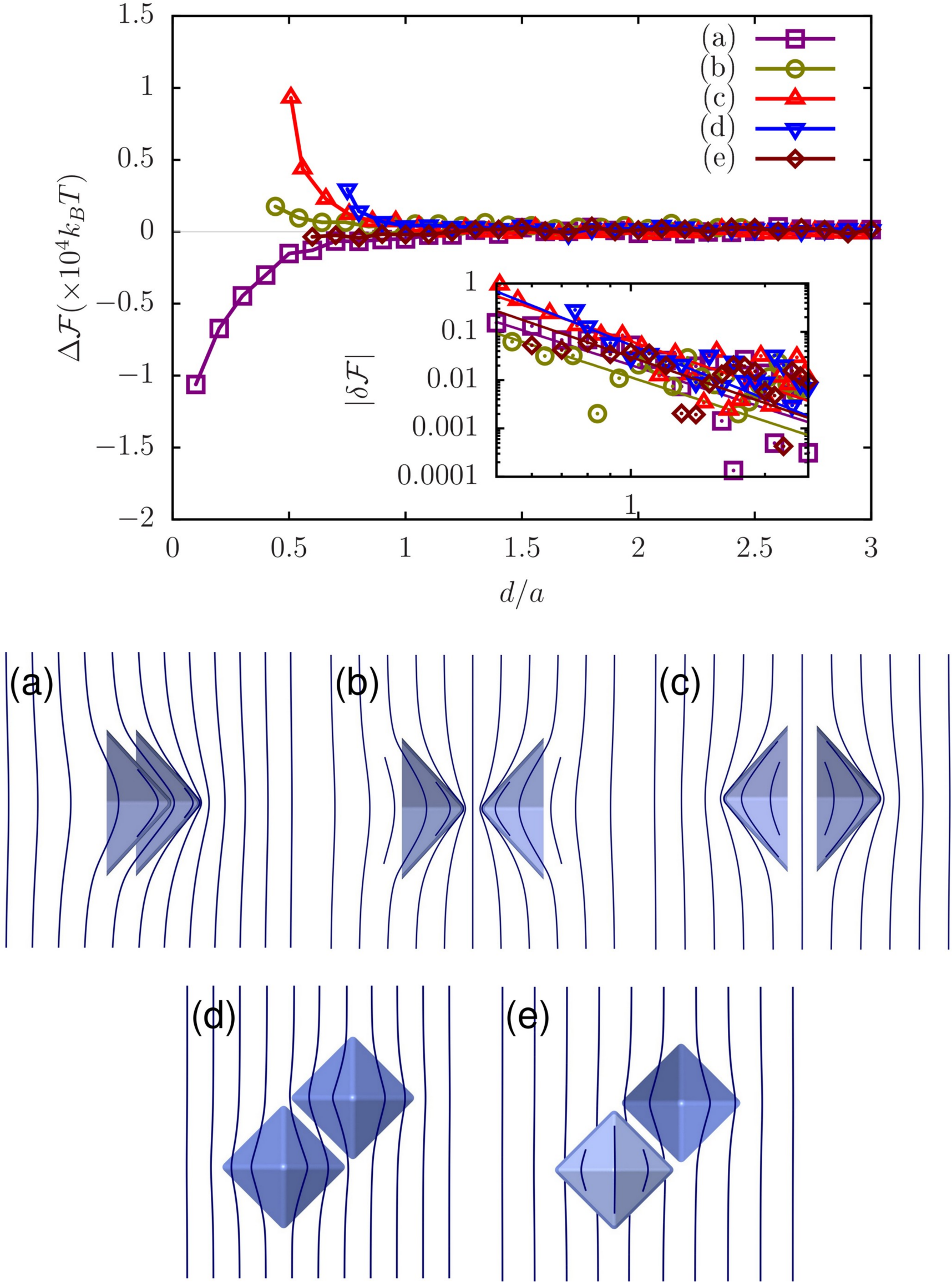}
		\caption{Interactions of two identical CPCs with $2\psi=95^\circ$ and $\vec{H}\perp\hat{n}_0$ in terms of particle-particle distances, $d$, where $\Delta\mathcal{F} = \mathcal{F} - 2\mathcal{F}_\text{iso}$ and $\mathcal{F}_\text{iso}(\theta=90^\circ, \phi=45^\circ)=1.732\times10^4k_\text{B}T$. The particle-particle arrangements, (a) coaxial and parallel ($\alpha=90^\circ$, $\theta_1=\theta_2 = 90^\circ$ and $\phi_1=\phi_2=45^\circ$), (b) coaxial and apex-apex antiparallel ($\alpha=90^\circ$, $\theta_1=90^\circ$, $\theta_2=-90^\circ$ and $\phi_1=\phi_2=45^\circ$), (c) coaxial and base-base antiparallel ($\alpha=90^\circ$, $\theta_1=-90^\circ$, $\theta_2 = 90^\circ$ and $\phi_1=\phi_2=45^\circ$), (d) noncoaxial and parallel ($\alpha=45^\circ$, $\theta_1=\theta_2 =90^\circ$, and $\phi_1=\phi_2=45^\circ$) and (e) noncoaxial and antiparallel ($\alpha=45^\circ$, $\theta_1=-90$, $\theta_2 =90^\circ$ and $\phi_1=\phi_2=45^\circ$). The inset log-log plot shows the fit of the numerical results by a power law function as $c_0 + c_1/(d/R)^c_2$ that $\delta\mathcal{F} = \mathcal{F} - c_0$. The unit of $c_0$ and $c_1$ is $10^4 k_B T$. The lines correspond to (a) $c_0=3.423\pm0.004$, $c_1=-0.020\pm0.001$ and $c_2=3.0\pm0.1$, (b) $c_0=3.445\pm0.003$, $c_1=0.056\pm0.001$ and $c_2=3.0\pm0.1$, (c) $c_0=3.407\pm0.003$, $c_1=0.064\pm0.003$ and $c_2=3.1\pm0.1$, (d) $c_0=3.431\pm0.004$, $c_1=0.053\pm0.006$ and $c_2=3.7\pm0.4$, and (e) $c_0=3.438\pm0.003$, $c_1=-0.030\pm0.006$ and $c_2=3.2\pm0.7$. The inter-particle distances are as in (a) $0.20a$, (b) $0.54a$, (c) $0.66a$, (d) $0.80a$, and (e) $0.80a$.}
		\label{fig07}
	\end{figure*}

	Similarly, figure \ref{fig06} shows the effective interaction potential between two CPCs with $\vec{H}\perp\hat{n}_0$ and $\phi=0^\circ$ in terms of inter-particle distances for the selected particle-particle arrangements as shown in Figs.~\ref{fig06}(a)-(g).In the absence of defects, the elastic deformations of the director and their energy cost are the main reason for inter-particle interactions.
	
	The CPCs attract each other in coaxial and parallel (Fig.~\ref{fig06}(a)), noncoaxial and antiparallel (Fig.~\ref{fig06}(e)), and non-coaxial and parallel (Fig.~\ref{fig06}(f)) configurations. They exhibit repulsive interactions in coaxial and apex-apex antiparallel (Fig.~\ref{fig06}(b)), coaxial and base-base antiparallel (Fig.~\ref{fig06}(c)), non-coaxial and parallel (Fig.~\ref{fig06}(d)), and non-coaxial and antiparallel (Fig.~\ref{fig06}(g)) configurations. Regardless of the distances between the particles, the repulsive and attractive interactions in Figs~\ref{fig06}(a)-\ref{fig06}(e) behave like the electrostatic dipole-dipole interaction, although there are no defects at the inner and outer tips. In Figs.\ref{fig05}(b)-(d) with repulsive interactions, the director shows a mirror symmetry about the plane passing through the bisector of the distances between two particles. The repulsive and attractive interactions in Figs~\ref{fig06}(f) and \ref{fig06}(g) respectively do not follow the nature of dipole-dipole interactions. Figs.~\ref{fig06}(d) and \ref{fig06}(f), and also Figs.~\ref{fig06}(e) and \ref{fig06}(g) show different repulsive and attractive interactions due to the spatial orientation of particles in nematic media. Fig.~\ref{fig06}(f) (in $\alpha=90^\circ$) is energetically an equilibrium arrangement compared to Fig.~\ref{fig06}(e)(in $\alpha=0^\circ$), and Fig.~\ref{fig06}(e) (in $\alpha=0^\circ$) indicates optimum structure compared to Fig.~\ref{fig06}(g) (in $\alpha=90^\circ$). We considered the $\alpha$ angle dependency on inter-particle interaction energies with more details in Supplemental Material Figs.~S2 and S3~\cite{Supplemental}. In non-coaxial and parallel (Fig.~\ref{fig06}(f)) configuration, the volume of elastic distortions between the particles decreases when particles approach each other with respect to large distances due to sharing of elastic distortions. In figure ~\ref{fig06}(e), the lateral triangular faces between two particles are antiparallel. Compared to the parallel structure, the antiparallel configuration contains a smaller practical volume between the faces, which may reduce elastic distortions. The inset log-log plot demonstrates that the long-range interactions are almost proportional to $(d/R)^{-3}$, consistent with experimental observation~\cite{Park.PhysRevLett.117.2016}. In agreement with the experimental observations~\cite{Park.PhysRevLett.117.2016}, the nesting of CPCs is an optimal structure in terms of energy, which reduces the elastic deformations between adjacent surfaces (see Fig.~\ref{fig06}(a)). The thermal fluctuations can also cause a transformation of the spatial arrangement of CPCs from the repulsive interaction in figure \ref{fig06}(b) to attractive interaction in Fig.~\ref{fig06}(e). Due to the repulsive interaction between the two CPC particles in Fig.~\ref{fig06}(c), the formation of an octahedron structure is also impossible.
	
	As shown in figure~\ref{fig04}, the energy of a single CPC increases slightly at $\theta=90^\circ$ for non-zero azimuthal angles. Here we study the behavior of CPCs in $\phi=45^\circ$ with $\vec{H}\perp\hat{n}_0$. Figure \ref{fig07} shows the effective energy potential between two CPCs with respect to the inter-particle distances for the selected particle-particle arrangements as shown in Figs.~\ref{fig07}(a)-(e). In the absence of defects, we can explain the repulsive and attractive interactions by imagining a dipole moment vector from the base to the apex for each CPC. They attract each other in coaxial and parallel (Fig.~\ref{fig07}(a)) and non-coaxial and antiparallel (Fig.~\ref{fig07}(e)) configurations.	In Figs.~\ref{fig07}(b) and (c), the director orientation has mirror-symmetry with respect to a plane parallel to the director and passing through the particle-particle separations. The particles exhibit repulsive interactions in coaxial and apex-apex antiparallel (Fig.~\ref{fig07}(b)), coaxial and base-base antiparallel (Fig.~\ref{fig07}(c)), and non-coaxial and parallel (Fig.~\ref{fig07}(d)) configurations. According to the inset log-log plot, the long-range interactions are nearly proportional to a dipolar potential in Figs.~\ref{fig07}(a)-(c). Coaxial configurations in $\phi=0^\circ$ are energetically favorable of $\phi=45^\circ$. From the fitted results in Figs.~\ref{fig07}(e) and (f), it can notice that long-range interactions show some deviations from the dipole-dipole interaction ($(d/R)^{-3})$. The nested configurations of CPCs are also energetically optimal structures. Due to the thermal fluctuations, the repulsive interaction in the apex-apex configuration can transform into a non-coaxial and antiparallel one with attractive interaction. The repulsive interaction between the CPCs in Fig.~\ref{fig07}(c) prevents the formation of octahedral shapes.

	\section{Summary}
	We have numerically investigated the spatial CPC arrangements as a thin shell with degenerate planar anchoring on inner and outer surfaces in the nematic host. The base-apex vector of CPC aligns along the far director alignment for $2\psi<95{^\circ}$ and perpendicularly for $2\psi>95^\circ$. The director shows the splay and the bend elastic distortions into the CPC with parallel and perpendicular orientations respectively. The elastic distortion changes the director orientation near the tip of the CPC and leads locally to a reduction of the scalar order parameter (or growth of the biaxiality). In the parallel case, the elastic distortions lead to two boojum defects at the inner and outer points of the pyramid tip. Our calculations show no defects in the perpendicular cases. The absence of boojums could be related to the atomic behavior of the sharp cones, which may  require a multi-scale approach with different levels of resolution and complexity. The nested structures can energetically form self-assembly of colloidal particles in agreement with experimental observations. The inter-particle interactions show long-range dipole-dipole behavior for both parallel and perpendicular cases. The apex-apex configurations can transform from repulsive to attractive interactions in the non-coaxial and antiparallel arrangements with small thermal fluctuations and spontaneous symmetry breaking. Our calculations show that the formation of octahedron structures is energetically impossible.  The interactions are independent of the CPCs azimuthal orientations for the parallel arrangements of the pyramids. However, the energies represent small differences in the azimuthal orientation for the perpendicular cases. According to the self-assembly of colloidal pyramidal cones with particular elastic distortions and defect arrangements, this study can lead to more attention for theoretical investigations in flat surfaces formed in a complex way in the NLCs.

\begin{acknowledgements}
	Authors acknowledge discussions with I. I. Smalyukh. 
\end{acknowledgements}

\bibliography{CPC}

\end{document}